# Out-of-plane Piezoelectricity and Ferroelectricity in Layered $\alpha$-In$_2$Se$_3$ Nano-flakes

Yu Zhou[1+], Di Wu[2+], Yihan Zhu[3], Yujin Cho[2], Qing He[4], Xiao Yang[1], Kevin Herrera[2], Zhaodong Chu[2], Yu Han[3], Michael C. Downer[2], Hailin Peng[1*] and Keji Lai[2*]

[1]Center for Nanochemistry, Beijing National Laboratory for Molecular Sciences (BNLMS), College of Chemistry and Molecular Engineering, Peking University, Beijing 100871, China

[2]Department of Physics, University of Texas at Austin, Austin TX 78712, USA

[3]Physical Sciences and Engineering Division, King Abdullah University of Science and Technology, Thuwal 23955-6900, Saudi Arabia

[4]Department of Physics, Durham University, Durham DH1 3LE, United Kingdom

**ABSTRACT:** Piezoelectric and ferroelectric properties in the two dimensional (2D) limit are highly desired for nanoelectronic, electromechanical, and optoelectronic applications. Here we report the first experimental evidence of out-of-plane piezoelectricity and ferroelectricity in van der Waals layered $\alpha$-In$_2$Se$_3$ nano-flakes. The non-centrosymmetric R3m symmetry of the $\alpha$-In$_2$Se$_3$ samples is confirmed by scanning transmission electron microscopy, second-harmonic generation, and Raman spectroscopy measurements. Domains with opposite polarizations are visualized by piezo-response force microscopy. Single-point poling experiments suggest that the polarization is potentially switchable for $\alpha$-In$_2$Se$_3$ nano-flakes with thicknesses down to ~ 10 nm. The piezotronic effect is demonstrated in two-terminal devices, where the Schottky barrier can be modulated by the strain-induced piezopotential. Our work on polar $\alpha$-In$_2$Se$_3$, one of the model 2D piezoelectrics and ferroelectrics with simple crystal structures, shows its great potential in electronic and photonic applications.

Two-dimensional (2D) van der Waals (vdW) materials encompassing a broad range of novel electronic,[1, 2] magnetic,[3] thermal,[4, 5] and optical properties[6, 7] have attracted substantial research interest over the past decade, promising the development of next-generation multi-functional devices. Among various functionalities, piezoelectricity and ferroelectricity widely exploited for the applications in memories,[8] capacitors,[9] actuators[10] and sensors[11] are relatively scarce in 2D materials. In 2H-stacking transitional metal dichalcogenides (TMDs) such as $MoS_2$, the inversion symmetry in bulk crystals is broken in ultrathin flakes with odd number of layers, leading to the in-plane piezoelectricity that has been theoretically predicted and experimentally demonstrated[12-14]. Recently, spontaneous in-plane polarization has also been reported in monolayer group IV monochalcogenides[15-17]. However, in device applications, out-of-plane piezoelectricity and ferroelectricity are more straightforward for circuit designs. To date, $CuInP_2S_6$ remains the only known vdW ferroelectric with out-of-plane polarization, although the crystal structure is rather complicated and the polarization is only switchable for films above 4 nm[18-21]. Since traditional ultrathin ferroelectric films such as $PbTiO_3$ and $BaTiO_3$ are plagued by dangling bonds and dead layers at the ferroelectric/metal interfaces [9], it is of great interest to explore new out-of-plane polarized 2D ferroelectrics for non-volatile memory and photovoltaic applications, as well as to enable 2D vdW heterostructures with novel functionalities [22-25].

2D out-of-plane ferroelectricity is highly nontrivial, as the depolarization field due to the lack of screening charges may strongly suppress spontaneous polarization in vdW materials. In a recent report, Ding et al. predicted that the layered semiconducting indium selenide (α-$In_2Se_3$) is a room-temperature out-of-plane polarized ferroelectric down to the single-layer limit (thickness ~ 1 nm), with a calculated electric dipole of 0.11 eÅ/unit cell[26]. $In_2Se_3$ has been widely explored for phase-change memory, thermoelectric, and photoelectric applications.[27, 28] Owing to its polymorphism and complicated phase diagram, however, even the crystal structure of the thermodynamically stable phase at the room temperature (commonly denoted as α-$In_2Se_3$) remains controversial.[29-31] Here we report the first experimental observation of out-of-plane piezoelectricity and ferroelectricity in multi-layer α-$In_2Se_3$. Using a combination of transmission electron microscopy (TEM), second-harmonic generation, Raman spectroscopy, and piezo-force microcopy (PFM), we show that our $In_2Se_3$ samples exhibit the rhombohedral R3m structure (Fig. 1a), which is non-centrosymmetric and supports the presence of a spontaneous polarization that is potentially switchable by an external bias. The piezotronic effect is demonstrated in that the charge transport in a prototypical device can be modulated by the piezoelectricity. Our work highlights the potential of 2D piezoelectric and ferroelectric materials for novel applications such as sensors, flexible electronics, and nano-electromechanical systems.

In this study, $In_2Se_3$ nano-flakes prepared by mechanical exfoliation onto conducting substrates and vapor-phase deposition (VPD) on flexible mica substrates are both studied[32]. Both types of samples have undergone a slow thermal annealing process before the characterizations (see Methods). In order to elucidate the out-of-plane structure of these nano-flakes, we use aberration corrected scanning transmission electron microscopy (AC-STEM) to directly image the cross sections of VPD-grown multi-layer nano-flakes. As shown in Fig. 1b, the cross-sectional samples are fabricated by focused-ion beam (FIB) cutting along the [120] and [100] axes of $In_2Se_3$ nano-flakes, respectively. From the annular bright-field (ABF) STEM image taken on the [120] cross-section (Fig.1c), the vdW gaps (in bright contrast) are clearly visible between the Se(1)-In(2)-Se(3)-In(4)-Se(5) quintuple layers. Interestingly, the ABF-STEM intensity profile in Fig. 1d indicates that the Se(3) atom is shifted off-center towards the neighboring In(2) atom, which breaks the inversion symmetry of each quintuple layer and gives rise to an out-of-plane dipole. This observation is consistent with the theoretical calculation[26]. The high angle annular dark field (HAADF) image and the ABF image along the [100] direction are shown in Supplementary Information Fig. S1. The STEM images and structural analysis suggest that the crystal structure of our $In_2Se_3$ nano-flakes follows the R3m symmetry[33, 34]. We note that α-$In_2Se_3$ samples in the R-3m or P6$_3$/mmc symmetry groups have also been reported in the literature [28, 30, 35]. While the origin of this discrepancy is not clear and may subject to future investigations, it is possible that the slow annealing pre-treatment is responsible for the polar structure observed in our samples.

The symmetry of our α-$In_2Se_3$ flakes is further explored by optical second-harmonic generation (SHG). Here a Ti: Sapphire femtosecond-pulsed laser with wavelength $\lambda_{ex}$ = 798 nm generates second-harmonic (SH) signals in reflection. Figure 2a compares the SH spectral intensity ($\lambda$ = 399 nm, all polarizations) generated by s-polarized incident laser from a vapor-phase deposited thin flake (thickness $t$ = 2 nm) with the 10-fold stronger SH peak that an identical pulse generates from an exfoliated thick flake ($t \sim$ 100 nm). These SH signals are, respectively, ~70 and ~350 times stronger than an identical pulse generates in reflection from a 2 nm thick GaAs film. Moreover, s-polarized SHG (Figure 2b), which has no contribution from the surface, is nearly as strong (averaged over azimuthal angles) as the p-polarized signal (Figure 2c), for which a surface contribution is allowed in principle. Note that VPD grown flakes with thickness from monolayer to four-layer all exhibit prominent SHG intensity (Fig. S3). These observations show that the SHG signal originates from the non-centrosymmetric bulk α-$In_2Se_3$ crystal, rather than from the broken inversion symmetry at the surface[36]. The result differs significantly from that of layered $MoS_2$, where SH intensities are negligible in even-layer and bulk samples due to the restoration of inversion symmetry[37, 38]. The azimuthal angle dependence of the SHG intensity is also measured on the VPD sample. For the R3m symmetry, SHG intensities take the form $I$ (s-in/s-out) = $I_0$ $\cos^2(3\theta)$ and $I$ (s-in/p-out) = $I_0$ $(A+B\cdot\cos(3\theta))^2$ in each polarization configuration. Here $\theta$ is the azimuthal angle from [120] direction, and $I_0$, A, and B are constants determined by Fresnel coefficients and the nonlinear susceptibility tensor. As shown in Figs. 2b and 2c, the calculated responses fit well to the s-(Figure 2b) and p-polarized (Figure 2c) SHG data in each configuration. In s-in/s-out configuration, only one component of the susceptibility tensor, $\chi_{yyy}$, generates the SH signal; in s-in/p-out configuration, the out-of-plane component, $\chi_{zyy}$, also contributes to the signal. The SHG data are therefore consistent with the conclusion that the symmetry group of our $\alpha$-In$_2$Se$_3$ crystals is R3m.

The broken inversion symmetry and polar structure in In$_2$Se$_3$ do not ensure its ferroelectricity, which necessarily requires the presence of a spontaneous polarization that is switchable under external electric fields. In order to investigate the piezoelectricity and ferroelectricity of the α-In$_2$Se$_3$ samples, PFM measurements (see Methods) have been carried out. Figure 3a shows the atomic force microscopy (AFM) image of a thick (> 100 nm) exfoliated In$_2$Se$_3$ flake with atomically smooth terraces. The out-of-plane PFM phase and amplitude images in Fig. 3b and 3c show two distinct regions with 180° phase difference, corresponding to domains with up and down polarization vectors perpendicular to the flake surface; whereas the domain walls appear as darker lines in the PFM amplitude image (Fig. 3c). Thinner flakes with thicknesses ranging from 3 nm to 60 nm (Fig. 3d) are also exfoliated onto gold substrates for PFM studies. As shown in Figs. 3e and 3f, clear out-of-plane domains can be observed. It is worth noting that some but not all of the domain walls coincide with the location of the flake edges, which suggests that the PFM phase contrasts are more likely coming from real polarization contribution rather than other artifacts between different layers. Unlike CuInP$_2$S$_6$, the In$_2$Se$_3$ flake does not display an obvious thickness dependence on the PFM amplitude contrast, which is consistent with the theoretical calculations[27]. To rule out the possibility that the PFM contrast is caused by the coexistence of different phases, local Raman spectroscopy is performed in this sample at different locations (marked with numerical labels in Fig. 3e), with the corresponding Raman spectra shown in Fig. 3g. Three prominent peaks, A(LO+TO) mode at 104 cm$^{-1}$ and A(LO) mode at 182 and 203 cm$^{-1}$, can be observed at locations 2-6. (Note that the regions with $t$ = 3 nm might have been oxidized by the Raman excitation laser.) The Raman frequencies are distinctly different from those of another room-temperature stable phase ($\beta$ phase), with A(LO+TO), A(TO), and A(LO) Raman modes centered at ~110cm$^{-1}$, ~175cm$^{-1}$ and ~205cm$^{-1}$, respectively[30, 39]. In accordance with previous Raman work,[31] the presence of the A(LO) mode indicates a lack of inversion symmetry in the R3m structure, consistent with the aforementioned STEM and SHG data.

We have also performed PFM tip poling experiments to study the ferroelectric hysteresis behavior of $\alpha$-In$_2$Se$_3$. Unfortunately, due to the small bulk resistivity, significant leakage current usually takes place before the switching events across the entire sample. Fig. S4 shows the I-V characteristics across a 15-nm-thick In$_2$Se$_3$ flake between a 1 μm × 1 μm Au pad and the bottom electrode. Substantial leakage current is observed for a bias beyond ±3 V, indicative of a large amount of defects (most likely Se vacancies) and charge carriers in the material. Because of the charge screening, we are not able to demonstrate the conventional remnant P-E hysteresis loop by the Sawyer-Tower method[40]. Nevertheless, we show that is it possible to obtain bias-on PFM hysteresis loops at individual points of the In$_2$Se$_3$ flakes. An example on a 20-nm-thick sample is seen in Figs. 4a and 4b, where a stiff cantilever with a spring constant of 40 N/m is used and the DC bias voltage is swept between -3 V and +6 V with an AC voltage of 800 mV. Here the amplitude response shows a butterfly loop with an opening of ~ 1.5 V, whereas the phase switches 180° at the same turning points. The unsaturated amplitude signal is likely due to the significant leakage (high concentration of free carriers) of the samples, although we cannot exclude the possibility of surface charging effect. The offset of the loop from zero bias is from the Schottky barrier difference between the upper ($In_2Se_3$ / IrPt tip) and lower ($In_2Se_3$ / Au substrate) surfaces of the sample. Similar results are acquired on flakes with thicknesses down to ~ 10 nm and the data resemble the hysteresis loops in standard ferroelectrics like PZT (Fig. S5). We emphasize that this extrinsic leakage effect may be mitigated by doping of the opposite type of charged impurities or using a different growth mechanism such as molecular-beam epitaxy (MBE). The same practice, for instance, has been successfully adopted to suppress bulk carriers in the $Bi_2Se_3$ family of topological insulators[41, 42].

Finally, a flexible $In_2Se_3$ device taking advantage of its out-of-plane piezoelectricity was demonstrated in Fig. 5. Standard photolithography is used to fabricate two-terminal devices on the VPD-grown multi-layer $In_2Se_3$ flakes (~10 nm) on mica substrates. High work-function metal Pd (20 nm) are deposited on the sample surface to form Schottky contacts. As shown in Fig. 5a, the source-drain current increases (decreases) considerably when the $In_2Se_3$ flake is under a small tensile (compressive) strain of ± 0.1%. Such a piezotronic effect has been previously reported in zinc oxide thin films[43], where the Schottky barrier height is modulated by the bound charges induced by the piezoelectricity of the semiconducting material, as schematically illustrated in Figs. 5b – 5e. Fifteen 2D $In_2Se_3$ devices have been investigated and all display the same transport characteristics in our experiment, suggesting a robust piezotronic effect that can be utilized for electromechanical energy transduction applications.

To summarize, the out-of-plane piezoelectricity and ferroelectricity in multi-layer α-$In_2Se_3$ are explored by a combination of structural, optical, and electrical characterizations. The non-centrosymmetric R3m crystal symmetry is confirmed by STEM, SHG, and Raman measurements. Ferroelectric domains are clearly visualized by PFM and the out-of-plane polarization is potentially switchable in samples with ~ 10 nm of thickness. Finally, the modulation of charge transport by bending of the substrate has also been demonstrated with a flexible device with mica substrate, showing great potential for applications in nanoscale electromechanical devices and piezotronic sensors. With further reduction of the bulk carrier density, it is possible that ferroelectric $In_2Se_3$ can be realized down to the single layer limit, which is highly desirable for memory, sensing, and photovoltaic applications.

Methods

*Sample preparation.* $In_2Se_3$ nano-flakes were grown on flexible fluorophlogopite mica substrates via vdW epitaxy in a pressure controllable vapor deposition system equipped with a 1-inch quartz tube.[32] The $In_2Se_3$ powder source (99.99%, Alfa Aesar) was heated to 690-750 ℃ at the center of the tube furnace. The vapor was transported downstream by ~50 sccm Ar gas with pressure controlled at ~50 Torr. The growth of $In_2Se_3$ nano-flakes occurred on the mica substrates placed 7-12 cm away from the heated center. After growth, the chamber was naturally cooled down to room temperature. Thin flakes were also exfoliated onto Au surface from the bulk taken out of the growth chamber for PFM measurements.

*STEM.* The cross-section TEM samples were prepared by focused ion beam cutting from two directions and were characterized by an aberration-corrected and monochromated G2 cubed Titan 60-300 electron microscope under 60 kV.

*SHG microscopy.* A continuous-wave Ti: Sapphire laser operating at ~798 nm, 76 MHz repetition rate, 150 fs pulse duration with *s*-polarization was focused on the sample surface at incident angle

θ = 45°, and SHG signals with *s* or *p*-polarization were collected in a reflection geometry. A photomultiplier tube (PMT) with bandpass filters was used to suppress the fundamental wave. It was confirmed that SHG signals scaled quadratically with the incident fundamental intensity.

*PFM measurements.* PFM measurements were conducted using a Park XE-70 system equipped with a Zurich HF2LI lock-in amplifier. P-E hysteresis loops were obtained with Asylum Research MFP-3D Infinity. Stiff cantilevers with a spring constant of 40 N/m were used to eliminate the electrostatic contribution.

*Raman spectroscopy characterization.* Raman spectroscopy was carried out using Witec Alpha 300 micro-Raman confocal microscope with a 488 nm laser excitation. The laser power was minimized to avoid burning the flakes.

*Device fabrication and measurements.* The flexible $In_2Se_3$ two-terminal devices were achieved by standard photolithography process, electron-beam deposition of Pd/Au (20 nm/50 nm) and gold wire pasted by silver epoxy for external connection. I-V measurements were conducted with a Keithley 4200 Semiconductor Characterization System.

ASSOCIATED CONTENT

**Supporting Information**

The HAADF TEM images, the setup of SHG, piezoelectricity measurements with devices are given in this section. This material is available free of charge via the Internet.

AUTHOR INFORMATION

**Notes**

[+]Y. Zhou and D. Wu contributed equally to this work.

The authors declare no competing financial interest.

**Corresponding Authors**

*Email: hlpeng@pku.edu.cn; kejilai@physics.utexas.edu

ACKNOWLEDGEMENTS

The PFM work is supported by the Welch Foundation Grant F-1814. D.W. and Z.C. also acknowledges the support from NSF EFRI under Award # EFMA-1542747. The SHG work (Y. C. and M. C. D.) is supported by Welch Grant F-1038. The sample synthesis and device fabrication work are supported by the National Basic Research Program of China (No. 2014CB932500) and the National Natural Science Foundation of China (No. 21525310).

**Reference**

(1) Castro Neto, A. H.; Guinea, F.; Peres, N. M. R.; Novoselov, K. S.; Geim, A. K. *Rev. Mod. Phys.* **2009,** 81, 109-162.

(2) Wang, Q. H.; Kalantar-Zadeh, K.; Kis, A.; Coleman, J. N.; Strano, M. S. *Nat. Nanotechnol.* **2012,** 7, 699-712.

(3) Huang, B.; Clark, G.; Navarro-Moratalla, E.; Klein, D. R.; Cheng, R.; Seyler, K. L.; Zhong, D.; Schmidgall, E.; McGuire, M. A.; Cobden, D. H. *arXiv preprint arXiv:1703.05892* **2017**.

(4) Wang, Y.; Xu, N.; Li, D.; Zhu, J. *Adv. Funct. Mater.* **2017**, 1604134.

(5) Zhou, Y.; Jang, H. J.; Woods, J. M.; Xie, Y. J.; Kumaravadivel, P.; Pan, G. A.; Liu, J. B.; Liu, Y. H.; Cahill, D. G.; Cha, J. J. *Adv. Funct. Mater.* **2017,** 27.

(6) Jones, A. M.; Yu, H. Y.; Ghimire, N. J.; Wu, S. F.; Aivazian, G.; Ross, J. S.; Zhao, B.; Yan, J. Q.; Mandrus, D. G.; Xiao, D.; Yao, W.; Xu, X. D. *Nat. Nanotechnol.* **2013,** 8, 634-638.

(7) Mak, K. F.; Shan, J. *Nat. Photonics* **2016,** 10, 216-226.

(8) Scott, J. F. *Ferroelectrics* **2000,** 236, 247-258.

(9) Stengel, M.; Spaldin, N. A. *Nature* **2006,** 443, 679-682.

(10) Crawley, E. F.; Deluis, J. *Aiaa J.* **1987,** 25, 1373-1385.

(11) Wang, X. D.; Zhou, J.; Song, J. H.; Liu, J.; Xu, N. S.; Wang, Z. L. *Nano Lett.* **2006,** 6, 2768-2772.

(12) Duerloo, K. A. N.; Ong, M. T.; Reed, E. J. *J. Phys. Chem. Lett.* **2012,** 3, 2871-2876.

(13) Wu, W. Z.; Wang, L.; Li, Y. L.; Zhang, F.; Lin, L.; Niu, S. M.; Chenet, D.; Zhang, X.; Hao, Y. F.; Heinz, T. F.; Hone, J.; Wang, Z. L. *Nature* **2014,** 514, 470-474.

(14) Zhu, H. Y.; Wang, Y.; Xiao, J.; Liu, M.; Xiong, S. M.; Wong, Z. J.; Ye, Z. L.; Ye, Y.; Yin, X. B.; Zhang, X. *Nat. Nanotechnol.* **2015,** 10, 151-155.

(15) Chang, K.; Liu, J. W.; Lin, H. C.; Wang, N.; Zhao, K.; Zhang, A. M.; Jin, F.; Zhong, Y.; Hu, X. P.; Duan, W. H.; Zhang, Q. M.; Fu, L.; Xue, Q. K.; Chen, X.; Ji, S. H. *Science* **2016,** 353, 274-278.

(16) Fei, R. X.; Kang, W.; Yang, L. *Phys. Rev. Lett.* **2016,** 117, 097601.

(17) Wang, H.; Qian, X. F. *2d Materials* **2017,** 4, 015042.

(18) Belianinov, A.; He, Q.; Dziaugys, A.; Maksymovych, P.; Eliseev, E.; Borisevich, A.; Morozovska, A.; Banys, J.; Vysochanskii, Y.; Kalinin, S. V. *Nano Lett.* **2015,** 15, 3808-3814.

(19) Susner, M. A.; Belianinov, A.; Borisevich, A.; He, Q.; Chyasnavichyus, M.; Demir, H.; Sholl, D. S.; Ganesh, P.; Abernathy, D. L.; McGuire, M. A.; Maksymovych, P. *Acs Nano* **2015,** 9, 12365-12373.

(20) Chyasnavichyus, M.; Susner, M. A.; Ievlev, A. V.; Eliseev, E. A.; Kalinin, S. V.; Balke, N.; Morozovska, A. N.; McGuire, M. A.; Maksymovych, P. *Appl. Phys. Lett.* **2016,** 109, 172901.

(21) Liu, F. C.; You, L.; Seyler, K. L.; Li, X. B.; Yu, P.; Lin, J. H.; Wang, X. W.; Zhou, J. D.; Wang, H.; He, H. Y.; Pantelides, S. T.; Zhou, W.; Sharma, P.; Xu, X. D.; Ajayan, P. M.; Wang, J. L.; Liu, Z. *Nat. Commun.* **2016,** 7, 12357.

(22) Butler, K. T.; Frost, J. M.; Walsh, A. *Energy & Environmental Science* **2015,** 8, 838-848.

(23) Morris, M. R.; Pendlebury, S. R.; Hong, J.; Dunn, S.; Durrant, J. R. *Adv. Mater.* **2016,** 28, 7123-7128.

(24) Geim, A. K.; Grigorieva, I. V. *Nature* **2013,** 499, 419-425.

(25) Grinberg, I.; West, D. V.; Torres, M.; Gou, G. Y.; Stein, D. M.; Wu, L. Y.; Chen, G. N.; Gallo, E. M.; Akbashev, A. R.; Davies, P. K.; Spanier, J. E.; Rappe, A. M. *Nature* **2013,** 503, 509-512.

(26) Ding, W.; Zhu, J.; Wang, Z.; Gao, Y.; Xiao, D.; Gu, Y.; Zhang, Z.; Zhu, W. *Nat. Commun.* **2017,** 8, 14956.

(27) Lee, H.; Kang, D. H.; Tran, L. *Mater. Sci. Eng. B-solid* **2005,** 119, 196-201.

(28) Han, G.; Chen, Z. G.; Drennan, J.; Zou, J. *Small* **2014,** 10, 2747-2765.

(29) Lai, K. J.; Peng, H. L.; Kundhikanjana, W.; Schoen, D. T.; Xie, C.; Meister, S.; Cui, Y.; Kelly, M. A.; Shen, Z. X. *Nano Lett.* **2009,** 9, 1265-1269.

(30) Tao, X.; Gu, Y. *Nano Lett.* **2013,** 13, 3501-3505.

(31) Lewandowska, R.; Bacewicz, R.; Filipowicz, J.; Paszkowicz, W. *Mater. Res. Bull.* **2001,** 36, 2577-2583.

(32) Lin, M.; Wu, D.; Zhou, Y.; Huang, W.; Jiang, W.; Zheng, W. S.; Zhao, S. L.; Jin, C. H.; Guo, Y. F.; Peng, H. L.; Liu, Z. F. *J. Am. Chem. Soc.* **2013,** 135, 13274-13277.

(33) Wu, D.; Pak, A. J.; Liu, Y. N.; Zhou, Y.; Wu, X. Y.; Zhu, Y. H.; Lin, M.; Han, Y.; Ren, Y.; Peng, H. L.; Tsai, Y. H.; Hwang, G. S.; Lai, K. J. *Nano Lett.* **2015,** 15, 8136-8140.

(34) Zhao, J. G.; Yang, L. X. *J. Phys. Chem. C* **2014,** 118, 5445-5452.

(35) Popovic, S.; Tonejc, A.; Grzetaplenkovic, B.; Celustka, B.; Trojko, R. *J. Appl. Crystallogr.* **1979,** 12, 416-420.

(36) Boyd, R. W., *Nonlinear optics*. 1992.

(37) Li, Y. L.; Rao, Y.; Mak, K. F.; You, Y. M.; Wang, S. Y.; Dean, C. R.; Heinz, T. F. *Nano Lett.* **2013,** 13, 3329-3333.

(38) Kumar, N.; Najmaei, S.; Cui, Q. N.; Ceballos, F.; Ajayan, P. M.; Lou, J.; Zhao, H. *Phys. Rev. B* **2013,** 87, 161403.

(39) Balakrishnan, N.; Staddon, C. R.; Smith, E. F.; Stec, J.; Gay, D.; Mudd, G. W.; Makarovsky, O.; Kudrynskyi, Z. R.; Kovalyuk, Z. D.; Eaves, L. *2D Mater* **2016,** 3, 025030.

(40) Dawber, M.; Rabe, K. M.; Scott, J. F. *Rev. Mod. Phys.* **2005,** 77, 1083-1130.

(41) Hong, S. S.; Cha, J. J.; Kong, D. S.; Cui, Y. *Nat. Commun.* **2012,** 3, 757.

(42) Chang, C. Z.; Zhang, J. S.; Feng, X.; Shen, J.; Zhang, Z. C.; Guo, M. H.; Li, K.; Ou, Y. B.; Wei, P.; Wang, L. L.; Ji, Z. Q.; Feng, Y.; Ji, S. H.; Chen, X.; Jia, J. F.; Dai, X.; Fang, Z.; Zhang, S. C.; He, K.; Wang, Y. Y.; Lu, L.; Ma, X. C.; Xue, Q. K. *Science* **2013,** 340, 167-170.

(43) Wen, X. N.; Wu, W. Z.; Ding, Y.; Wang, Z. L. *Adv. Mater.* **2013,** 25, 3371-3379.

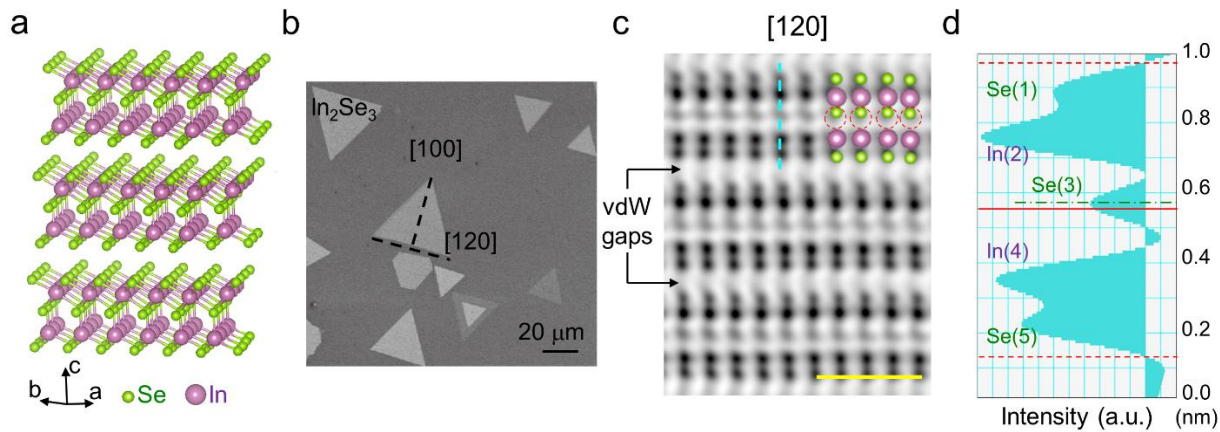

**Figure 1. Atomic structure of layered In$_2$Se$_3$ nano-flake.** (**a**) Crystal structure of $\alpha$-In$_2$Se$_3$ in space group of R3m. (**b**) Scanning electron micrograph (SEM) of VPD grown In$_2$Se$_3$ flakes. The [120] and [100] zone axes, along which the flakes are cut for STEM studies, are labeled in the image. (**c**) Cross-sectional annular bright-field (ABF) STEM image of an In$_2$Se$_3$ flake cut along the [120] direction. Se and In atoms, as well as the van der Waals gaps, are indicated in the figure. The scale bar is 1 nm. (**d**) Intensity profile along the blue dashed line in (c). The center of the Se(3) atom is slightly shifted with respect to the central position of the quintuple layer.

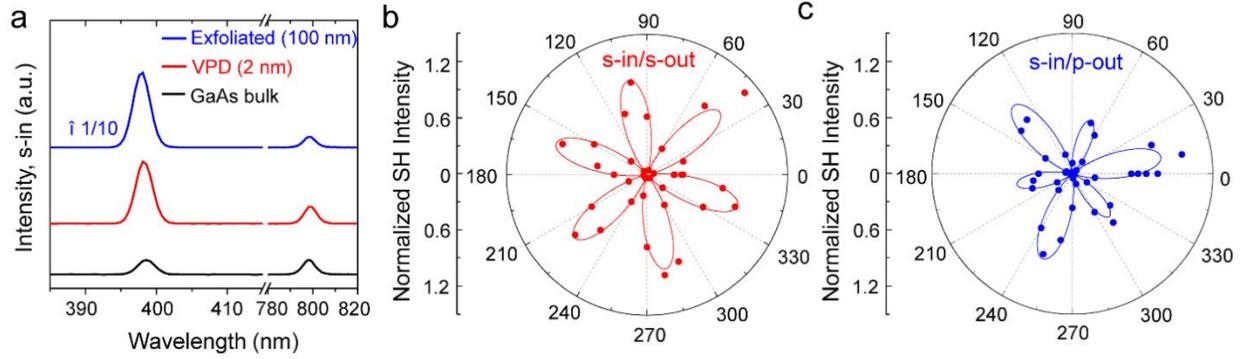

**Figure 2. SHG of In$_2$Se$_3$ nano-flakes.** (**a**) SH spectral intensity at 399 nm (all polarizations) generated in reflection from VPD grown (red), exfoliated (blue) In$_2$Se$_3$ flakes (~100 nm thick) and GaAs bulk. All samples are excited by equally intense s-polarized laser pulses with a wavelength of $\lambda_{ex}$ = 798 nm, which is strongly suppressed in the plot with respect to the corresponding SH peak. (**b**) Dependence of SHG signals on sample azimuthal angle $\theta$ in the s-in/s-out configuration. The red curve is a fit to $I = I_0 \cos^2(3\theta)$. (**c**) $\theta$-dependence of SHG signals in the s-in/p-out configuration. The blue curve is a fit to $I = I_0 (A+B\cdot\cos(3\theta))^2$. The left-hand axes in (b) and (c) denote the radial scale.

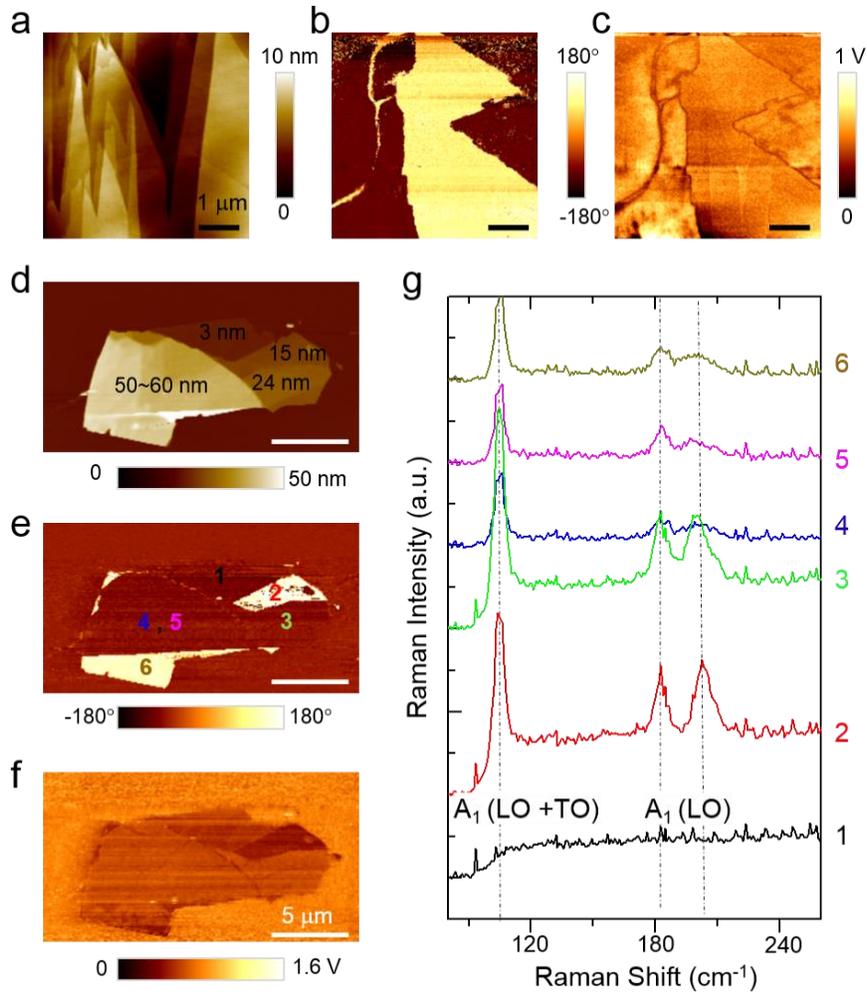

**Figure 3| Ferroelectric domains of α-In$_2$Se$_3$ flakes.** (**a - c**) AFM, PFM phase and amplitude images of a thin α-In$_2$Se$_3$ flake (> 100 nm). (**d - f**) AFM, PFM phase and amplitude images of a thin α-In$_2$Se$_3$ flake exfoliated onto gold surface. The PFM phase contrast of 180° in both samples indicates the presence of different domains with opposite out-of-plane polarizations. The scale bars are 1 μm in (a – c) and 5 μm in (d – f). (**g**) Raman spectra at different locations labeled in (d). The prominent Raman peaks at 104 cm$^{-1}$ and 182 / 203 cm$^{-1}$ are associated with A(LO+TO) and A(LO) modes, respectively.

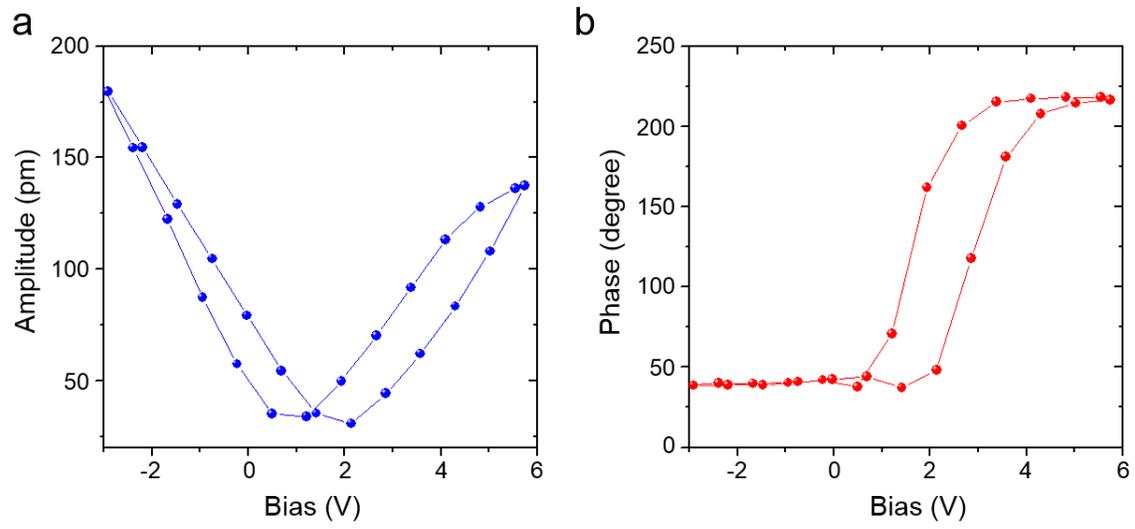

**Figure 4| Polarization reversal under external electrical field.** On-field **(a)** PFM amplitude and **(b)** PFM phase hysteresis loops on a 20-nm-thick flake.

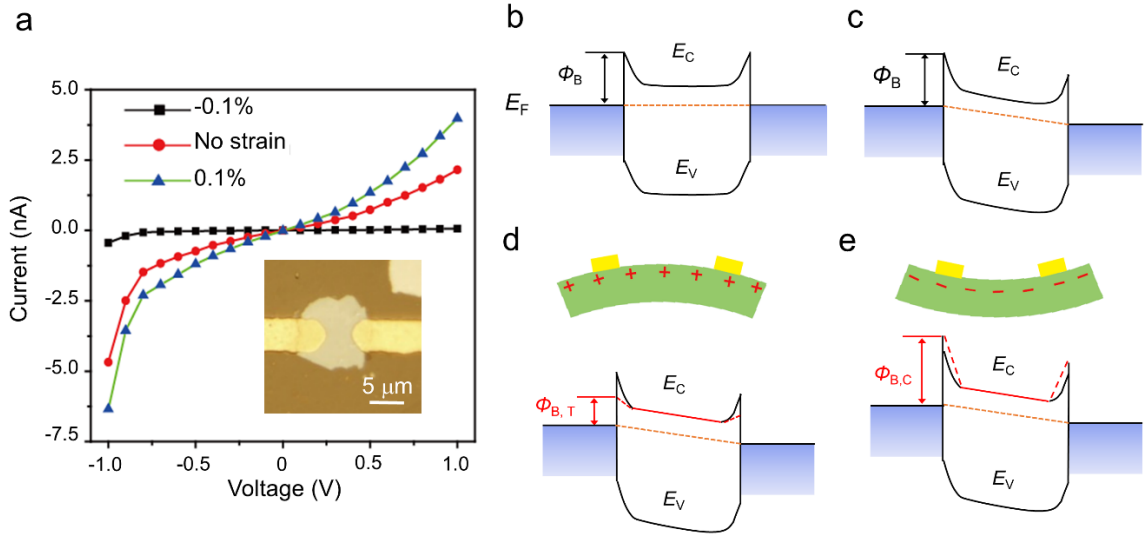

**Figure 5. Piezotronic effect on prototypical In$_2$Se$_3$ devices.** (**a**) Typical I-V characteristics of two-terminal In$_2$Se$_3$ devices under compressive (black), zero (red), and tensile (blue) strains by bending the substrate. Inset shows an image of typical 2D In$_2$Se$_3$ device. (**b - e**) Band diagrams under different conditions: (b) Zero strain with zero source-drain bias. (c) Zero strain with non-zero source-drain bias. (d) Tensile strain with non-zero source-drain bias, resulting in lower Schottky barriers ($\Phi_{B,T}$) and enhanced current. (e) Compressive strain with non-zero source-drain bias, resulting in higher Schottky barriers ($\Phi_{B,C}$) and reduced current.